\documentstyle[preprint,aps]{revtex}
\tightenlines

\begin{document}
\title{Comment on ``A proposed method for measuring the electric dipole moment of
the neutron using acceleration in an electric field gradient and ultracold
neutron interferometry '', II}
\author{S. K. Lamoreaux* and R.Golub$^{\#}$}
\address{* Los Alamos National Laboratory, Physics Div., M.S. H803,\\
Los Alamos, NM 87545, USA\\
$^{\#}$ Hahn Meitner Institut, Glienicker Str. 100, D-14109, Berlin, Germany}
\date{\today}
\maketitle

\begin{abstract}
We discuss the proposal of Freedman, Ringo and Dombeck \cite{ringo} to
search for the neutron electric dipole moment by use of the acceleration of
ultracold neutrons in an inhomogeneous electric field followed by
amplification of the resulting displacement by several methods involving
spin independent interactions (gravity) or reflection from curved (spin
independent) mirrors. We show that the proposed technique is inferior to the
usual methods based on magnetic resonance.
\end{abstract}


\section{Introduction}

Since the appearance of our original paper \cite{NIM} there has been some
additional discussion \cite{peshkin} and \cite{NIMDR}, so that it now seems
reasonable to publish a revised version with a more detailed exposition of
our quantum mechanical treatment (sec. II B.)

Searches for a neutron (or elementary particle) electric dipole moment (edm)
are interesting because an observation of an edm would be a demonstration of
the violation of time reversal (T) invariance outside the $K^{0}$ system.
Freedman et al. \cite{ringo} have proposed a new method to search for the
neutron edm. The method is claimed to offer the possibility of vastly
improved sensitivity due to the amplification of the effects of the
interaction of an edm with an electric field by means of subsequent motion
in a gravitational field or reflection from a convex mirror. In the present
paper we will review this proposal and show that the claimed gain in
sensitivity is based on a misunderstanding of a semi-classical model of the
processes involved.

According to the proposal, Ultra-cold neutrons (UCN) polarized along the $x$
axis enter a chamber (accelerator) where the electric field directed along $%
z $ has a gradient in the $x$ direction. Since the incident spin state which
is an eigenstate of $\sigma _{x}$ can be considered as a coherent
superposition of the two eigenstates $\left| \pm \right\rangle _{z}$ of $%
\sigma _{z}$, we can suppose that neutrons possessing a non-zero edm in one
of the eigenstates will suffer an acceleration 
\begin{equation}
a_{x}=\pm \frac{\mu _{e}}{m}\left( \frac{\partial E_{z}}{\partial x}\right).
\label{one}
\end{equation}
After time $T_{1}$ the two spin states will be separated by a distance 
\begin{equation}
\delta x=\frac{\mu _{e}}{m}\left( \frac{\partial E_{z}}{\partial x}\right)
T_{1}^{2}  \label{two}
\end{equation}
After the period of acceleration, the neutrons are allowed to leave the
chamber in a vertical direction $\left( +z\right) $ rising against gravity.
After reflection from a surface inclined at 45$^{0},$ the horizontal
separation $\delta x$ is converted into a vertical separation $\delta
x=\delta z$. The neutrons then follow parabolic trajectories under the
influence of gravity. Because of the difference in initial heights, $\delta
z $, hence a difference in kinetic energy, the two spins will accumulate a
phase difference along the two trajectories 
\begin{equation}
\Delta \phi =\frac{1}{\hbar }\int \Delta Vdt=mg\Delta zT_{2}  \label{three}
\end{equation}
where $T_{2}$ is the time of flight along the parabolic trajector(ies). This
is called a 'gravitational amplifier' by the authors \cite{ringo}. It is
then proposed to measure this amplified phase difference between the states $%
\left| \pm \right\rangle _{z}$ as a precession of the polarization vector in
the $x,y$ plane: 
\begin{equation}
\Delta \phi =gT_{2}\mu _{e}\frac{\partial E_{z}}{\partial x}T_{1}^{2}
\label{four}
\end{equation}

Putting in practical values for the parameters the authors expect a
sensitivity to the edm of $10^{-28}\ e-cm$, which is superior to that
expected for other methods \cite{LG}. It is seen that the proposed effect
depends crucially on the description of the spin 1/2 system where the phase
difference is calculated along trajectories ending at different points while
the polarization is calculated by assuming the two states combine coherently
at a single point.

The authors also propose a second type of amplifier based on repeated
reflections from a curved surface, given by $Z(x)$. Then the two `points'
representing the two spin states separated by $\delta x$ due to the edm
acceleration $\left( \ref{one}\right) $ will reflect from portions of the
surface with slightly different slopes, differing by: 
\begin{equation}
\alpha \left( x\right) =\frac{\partial ^{2}Z\left( x\right) }{\partial x^{2}}%
\delta x  \label{five}
\end{equation}
and, after reflection the angle between the trajectories will increase by $%
2\alpha \left( x\right) $. Again, the increased separation is supposed to
result in an amplification of the phase difference and hence of the
precession in the $x,y$ plane. This has the same feature as the
``gravitational amplifier'': trajectories ending at increasingly distant
points are used to calculate a phase difference which is then thought to be
measured as a precession of the polarization which is calculated by
considering that the two states combine coherently at a single point.

\section{Discussion of the proposal according to different models}

As the present proposal is based on a mixing of models we will present a
discussion of the proposal from several view points. It is important that
the models be well defined and applied in a consistent manner. Mixing
different models leads to errors and confusion.

A detailed discussion of the relation between classical, semi-classical and
quantum mechanical descriptions of similar situations has been given in \cite
{stnse} with a more complete quantum mechanical discussion in \cite{plane}.
We will review these ideas with emphasis on the application to the present
model.

\subsection{Classical Model}

As an edm interacting with an electric field behaves identically to a
magnetic moment interacting with a magnetic field we choose to discuss the
problem in terms of magnetic moments in a constant magnetic field as this is
probably more familiar. We will see later that the introduction of field
gradients in the author's proposal is neither essential nor desirable.

A classical magnetic moment $\overrightarrow{\mu }$ coupled to an angular
moment $\overrightarrow{j}$, $\overrightarrow{\mu }=\gamma \overrightarrow{j}
$ in an external field $\overrightarrow{B}$, (of course the same discussion
will apply to an edm in an electric field) obeys the equation of motion 
\begin{equation}
\frac{d\overrightarrow{j}}{dt}=\overrightarrow{\mu }\times \overrightarrow{B}
\label{six}
\end{equation}
whose solution is seen to be a precession of $\overrightarrow{j}$ around $%
\overrightarrow{B}$ with the Larmor frequency $\omega _{L}=\gamma B$
independent of the angle between $\overrightarrow{j}$ and $\overrightarrow{B}
$ and which is a constant of the motion. If the field exists in a region of
length $L$, then the particle will cross the field in a time $t=L/v$ during
which time the spin will precess through an angle 
\begin{equation}
\varphi _{L}=\omega _{L}t=\omega _{L}L/v=\gamma BL/v  \label{seven}
\end{equation}

The spin components, $\sigma _{x,y}$, averaged over the velocity spectrum $%
f\left( v\right) $ of the beam,\bigskip\ will then be given by 
\begin{eqnarray}
\left\langle \sigma _{x}\right\rangle &=&\int dvf\left( v\right) \cos \left( 
\frac{\omega _{L}L}{v}\right)  \label{eight} \\
\left\langle \sigma _{y}\right\rangle &=&\int dvf\left( v\right) \sin \left( 
\frac{\omega _{L}L}{v}\right)  \nonumber
\end{eqnarray}
This is the basis of a technique that is often used to measure velocity
distributions.

Note that as $\omega _{L}L$ increases enough the polarization $\left\langle
\left| \sigma \right| \right\rangle $ is expected to decrease. In a
classical model the separation of states in space does not appear. As the
gravitational interaction is spin-independent there is no gravitational
amplification in this model. Moments entering the field directed
perpendicular to it undergo no energy change since $\overrightarrow{\mu }%
\cdot \overrightarrow{B}=0$.

The classical model is expected to give accurate results so long as the
separation between trajectories associated with different states is small
compared to the correlation (coherence) lengths of the wave function. Thus
the classical model breaks down in the Stern-Gerlach effect but gives a very
good description of Larmor precession. \cite{plane}. Therefore one could
conclude on this basis that the proposed amplification does not exist but we
are aware that this argument would not be seen as compelling.

\subsection{Quantum mechanical treatment}

Since the semi-classical model is in some ways the most difficult, primarily
because it is prone to misinterpretation, we will first discuss the quantum
mechanical model. Note that to our knowledge no quantum mechanical
demonstration of the alleged amplification effect has been presented, \cite
{ringo}, \cite{peshkin}, \cite{NIMDR}.\ 

We start with some general, elementary remarks. A wave function $\psi \left(
x\right) $ and its Fourier transform $\Psi \left( k\right) $ are
respectively, the probability amplitudes for finding the particle in a given
region of position or momentum space. Then a displacement of the particle is
equivalent to a phase shift (linear in momentum) of the momentum wave
function:

\begin{equation}
\psi \left( \overrightarrow{x}+\overrightarrow{\delta x}\right) =\int
d^{3}ke^{i\overrightarrow{k}\cdot \overrightarrow{x}}\Psi \left( 
\overrightarrow{k}\right) e^{i\overrightarrow{k}\cdot \left( \overrightarrow{%
\delta x}\right) }  \label{nine}
\end{equation}

A more general phase shift 
\begin{equation}
\Psi \left( \overrightarrow{k}\right) =\left| \Psi \left( \overrightarrow{k}%
\right) \right| e^{i\varphi \left( \overrightarrow{k}\right) }  \label{ten}
\end{equation}
can be considered as leading to a displacement, $\overrightarrow{\delta x}$,
such that 
\begin{equation}
\varphi \left( \overrightarrow{k}\right) =\varphi \left( \overrightarrow{k}%
_{o}\right) +\overrightarrow{\kappa }\cdot \overrightarrow{\nabla }%
_{k}\varphi \left( \overrightarrow{k}\right) =\varphi \left( \overrightarrow{%
k}_{o}\right) +\overrightarrow{\kappa }\cdot \overrightarrow{\delta x}
\label{new}
\end{equation}
only under the condition that $\Psi \left( \overrightarrow{k}\right) $ is
narrow enough that higher order terms in the expansion of $\varphi \left( 
\overrightarrow{k}\right) $ can be neglected. Also $\varphi \left( 
\overrightarrow{k}_{o}\right) $ should not play an important role. In the
case of neutron spin echo \cite{stnse} (or an edm accelerator with constant
electric field) \ where $\varphi \propto 1/k$ the condition for this is ($%
\kappa /k_{o}\ll 1$) where $\kappa $ is the width of the wave in \ momentum
space centered around $k_{o}$. In the spin echo case we also have $\delta
x\gg 1/\kappa $ so that the concept of a displacement between the states has
physical meaning \ and, as will be seen below, no polarization can be
observed under these conditions. It is necessary to cancel the phase shift
(this is called obtaining the echo), and thus eliminate the displacement $%
\delta $, in order to observe the polarization.

The reason for belaboring these rather self-evident points is that in the
work in question there is a tendency to make a distinction between the
displacement and the precession methods of searching for an edm whereas we
have seen that the two concepts are only different ways of looking at the
same phenomenon. There is only one phase for each quantum state and attempts
to separate this phase into various components can often lead to confusion.
This applies to the present case where a term linear in $k$ is singled out
as a `displacement' or in attempts at separating the phase into 'geometric'
and `dynamic' parts. This latter separation is strongly dependent on the
coordinate system used for the calculation - only the total phase remains
unchanged \cite{berry}.

If we consider a wave function $\Psi \left( \overrightarrow{k}\right) $ for
a state where the spins are initially polarized in the $x$ direction 
\begin{equation}
\left| \Psi \left( t=0\right) \right\rangle =\Psi \left( \overrightarrow{k}%
\right) \left( 
\begin{array}{c}
1 \\ 
1
\end{array}
\right)   \label{elf}
\end{equation}
and, as a result of a spin dependent interaction, the spin states pick up an
additional phase $\varphi _{\pm }$ so that at some time $t$ we have 
\begin{equation}
\left| \Psi \left( \overrightarrow{k},t\right) \right\rangle =\Psi \left( 
\overrightarrow{k}\right) \left( 
\begin{array}{c}
e^{i\varphi _{k}^{+}\left( t\right) } \\ 
e^{i\varphi _{k}^{-}\left( t\right) }
\end{array}
\right)   \label{new2}
\end{equation}
then the expectation value of $\overrightarrow{\sigma }$ will be 
\begin{equation}
\left\langle \sigma _{x}+i\sigma _{y}\right\rangle =\left\langle \Psi \left(
t\right) \left| \sigma _{x}+i\sigma _{y}\right| \Psi \left( t\right)
\right\rangle =\int \left| \Psi \left( \overrightarrow{k}\right) \right|
^{2}e^{i\left( \varphi _{k}^{+}-\varphi _{k}^{-}\right) }d^{3}k
\label{twelve}
\end{equation}

In the case where the spin dependent interaction, $V_{\pm }=\pm V_{o}$, is
small compared to the kinetic energy of the beam particles (this is
certainly the case for any interaction involving a particle edm and holds
for the spin echo case as well \cite{plane}) and slowly varying we can use
the WKB approximation to write

\begin{equation}
\varphi _{\pm }=\int dx\sqrt{k^{2}-2mV_{\pm }/\hbar ^{2}}\approx kx\mp \int
dx\frac{mV_{o}\left( x\right) }{\hbar ^{2}k}  \label{13}
\end{equation}
For $V_{o}\left( x\right) =const=\mu B$ or $\mu _{e}E$ we can write, taking$%
\ \alpha =\omega _{L}Lm/2\hbar $, with $\omega _{L}=2\mu B/\hbar $ or $2\mu
_{e}E/\hbar $  
\begin{equation}
\varphi _{k}^{\pm }=\pm \frac{\alpha }{k}=\pm \left( \varphi _{o}-\kappa
\delta x\right)   \label{aa2}
\end{equation}
where we have expanded around $k_{o}$, $\left( k=k_{o}+\kappa \right) $ for
a narrow spectrum centered on $k_{o}$ using $\delta x=\alpha /k_{o}^{2},$
see (\ref{new}), ($\varphi _{o}=\alpha /k_{o})$. The case where $V_{o}\left(
x\right) $ has a constant gradient, $V_{o}=\mu _{e}x\left( \frac{\partial E}{%
\partial x}\right) $ is seen not to introduce any significant differences
into (\ref{13}). Then

\begin{eqnarray}
\left\langle \sigma _{x}+i\sigma _{y}\right\rangle  &=&\left\langle \psi
^{\ast }\left( x-\delta x,t\right) \psi \left( x+\delta x,t\right)
e^{i2\varphi _{o}}\right\rangle   \label{16} \\
&=&\left\langle \psi _{+}^{\ast }\left( x,t\right) \psi _{-}\left(
x,t\right) \right\rangle \equiv I(t)  \label{new5}
\end{eqnarray}
where $\psi \left( x\right) $ is the Fourier transform of $\Psi \left(
k\right) $ and equ. (\ref{16}) is the result of the Wiener-Khintchin theorem
applied to (\ref{twelve}). We see that $\left\langle \sigma _{x}+i\sigma
_{y}\right\rangle $ in equ. (\ref{twelve}) as a function of $\varphi $
represents the Fourier transform of the beam velocity spectrum in agreement
with the classical calculation, equ. (\ref{eight}). From equ. (\ref{16}) it
follows that this is a decreasing function of $\varphi _{k}=$  $\varphi
_{k}^{+}-\varphi _{k}^{-}$ or $\delta x$ for large enough $\varphi _{k}$ so
that increasing $\varphi _{k}$ or $\delta x$ far enough will result in a
reduction of the net polarization. As $\delta x$ increases so much that the
correlation function in (\ref{16}) approaches zero we approach the case of
the Stern-Gerlach effect where the spin states can be considered as truly
separated and one can talk about a displacement without any confusion. Thus
we have seen that the discussion can be carried out equally either in terms
of $\varphi _{k}$ or $\delta x$, they both represent the same physical
situation.

The main thrust of the proposal under discussion \cite{ringo} is that after
a period in the electric field region the $\ x$ velocity of the UCN is
changed into the $z$ direction by reflection from a mirror at $45^{o}$.
Subsequent motion in the gravitational field is supposed to significantly
increase the phase difference $\varphi $, according to the (incorrect)
semi-classical argument presented in \cite{ringo} and outlined in the
introduction.

The key point is that no spin-independent interaction can influence the
phase shift between the two spin states. At the entrance to the
gravitational \ field drift region (called the ''amplifier'' by the authors)
the wave function will be of the form given by (\ref{new2}) and (\ref{aa2});
taking the Fourier transform

\begin{equation}
\left| \psi \left( \overrightarrow{x},t\right) \right\rangle ^{o}=\left( 
\begin{array}{c}
e^{i\varphi _{o}}\psi ^{o}\left( \overrightarrow{x}+\overrightarrow{\delta x}%
,t\right)  \\ 
e^{-i\varphi _{o}}\psi ^{o}\left( \overrightarrow{x}-\overrightarrow{\delta x%
},t\right) 
\end{array}
\right) =\left( 
\begin{array}{c}
\psi _{+}^{o}\left( \overrightarrow{x},t\right)  \\ 
\psi _{-}^{o}\left( \overrightarrow{x},t\right) 
\end{array}
\right)   \label{new3}
\end{equation}
for a narrow spectrum. The superscript $^{o\text{ }}$refers to the incoming
beam at the entrance orifice of the amplifier region.

We now calculate the wave function at the exit of this region. It is
sufficient to consider only the motion in the $z$ direction and confine
ourselves to the steady state situation. Then using the WKB method for the
case of the slowly varying gravitational potential, which is not necessarily
small compared to the particle kinetic energy, we can write the wave
function at the output (in the case of zero electric field in the
accelerator) 
\begin{equation}
\psi \left( z\right) =\sum_{k}A\left( k\right) e^{i\int_{0}^{z}\sqrt{\left(
k^{2}+K_{z}^{2}\right) }dz}=\sum_{k}A\left( k\right) e^{i\frac{\hbar ^{2}}{%
3m^{2}g}\left( k^{2}+K_{z}^{2}\right) ^{3/2}}e^{-i\frac{\hbar ^{2}}{3m^{2}g}%
k^{3}}  \label{aaa2}
\end{equation}
where $K_{z}^{2}\equiv 2m^{2}gz/\hbar ^{2}=m^{2}v_{z}^{2}/\hbar ^{2}$. $v_{z}
$ is the velocity of a particle that falls a distance $z$, starting from
rest. In order to investigate the relation between phase shifts and
displacements we wish to study the behavior of $\psi $ in a small region
located around $z=Z$, so we set $z=Z+\epsilon $, $\epsilon \ll Z$ and expand
the exponential in (\ref{aaa2}) around $Z$. 
\begin{equation}
\psi \left( z=Z+\epsilon \right) =\sum_{k}A\left( k\right) \exp i\left[ 
\frac{\hbar ^{2}}{3m^{2}g}\left( k^{2}+K_{Z}^{2}\right) ^{3/2}+\left(
k^{2}+K_{Z}^{2}\right) ^{1/2}\epsilon \right] e^{-i\frac{\hbar ^{2}}{3m^{2}g}%
k^{3}}  \label{aaa3}
\end{equation}

Since $A\left( k\right) $ represents a fairly narrow wave packet centered
around $k_{o}$ we can write $k=k_{o}+\kappa $, $\kappa \ll k_{o}\ll K_{Z}$
(the latter condition is necessary for the 'amplification' to be
significant, see below) and then we obtain from (\ref{aaa3}), \ keeping only
the terms in $\kappa $ and $\epsilon $ 
\begin{equation}
\psi \left( z=Z+\epsilon \right) \approx \sum_{\kappa }A\left( k_{o}+\kappa
\right) \exp i\left[ \frac{\hbar ^{2}k_{o}}{m^{2}g}\left( K_{Z}-k_{o}\right)
\kappa +K_{Z}\epsilon +\frac{k_{o}^{2}}{2K_{Z}}\epsilon +\frac{k_{o}}{K_{Z}}%
\kappa \epsilon \right]   \label{aaa6}
\end{equation}
The terms in (\ref{aaa6}) can be interpreted as follows: The first term
represents a displacement in position by $v_{o}t_{Z}$, the distance the
particle with the initial velocity, $v_{o}$, would go in the time $t_{Z}$
that the particle takes to fall to $Z$ and an additional displacement $%
E_{o}/mg$ associated with the boundary condition at $z=0$. The next two
terms represent the change in wavelength due to the acceleration during the
fall. Since $\sum_{\kappa }A\left( k_{o}+\kappa \right) e^{i\kappa \epsilon }
$ represents the envelope of the wave function at $z=0$, the last term in (%
\ref{aaa6}) shows that the envelope is spread by the factor $\eta
=K_{Z}/k_{o}\gg 1$ due to the energy dependence of the index of refraction
(dispersion) for the Schr\"{o}dinger wave.

When the electric field is switched on we have (using equ. (\ref{aa2}).) 
\begin{eqnarray}
\psi _{\pm }\left( z=Z+\epsilon \right)  &\approx &\sum_{\kappa }A\left(
k_{o}+\kappa \right) e^{i\frac{k_{o}}{K_{Z}}\kappa \epsilon }e^{i\varphi
_{k}^{\pm }}  \label{aaa5} \\
&\approx &\sum_{\kappa }A\left( k_{o}+\kappa \right) e^{i\frac{k_{o}}{K_{Z}}%
\kappa \epsilon }e^{\pm i\left( \varphi _{o}-\kappa \delta x\right) } 
\nonumber
\end{eqnarray}
displaying only the relevant terms. Now if we calculate the center of the
wave packet, $\epsilon _{o}$, according to $\left. \partial \varphi
/\partial \kappa \right| _{\epsilon _{o}}=0$, we find the center of the
packet is indeed shifted by 
\begin{equation}
\delta \epsilon =\eta \delta x  \label{aa1}
\end{equation}
in agreement with the ideas of ref. \cite{ringo}. The wave packet spreading
by the factor $\eta $, effects the electric field induced displacement by
the same factor during the transit through the amplifier region. This is the
quantum equivalent of the semi-classical argument given in \cite{ringo}.
However we see immediately from (\ref{aaa5}) that the $\kappa \delta x$ term
in the phase shift is a small part of the total edm induced, spin dependent
phase shift and neither term in the phase shift is altered by travel through
the 'amplifier' region. Thus from equ. (\ref{aaa5}) we have 
\begin{equation}
\left. \psi _{+}^{\ast }\psi _{-}\right| _{\left( z=Z+\epsilon \right)
}\approx \sum_{\kappa ,\kappa ^{\prime }}A^{\ast }\left( k_{o}+\kappa
^{\prime }\right) A\left( k_{o}+\kappa \right) e^{i\frac{k_{o}}{K_{Z}}\left(
\kappa -\kappa ^{\prime }\right) \epsilon }e^{i\left( 2\varphi _{o}-\left(
\kappa -\kappa ^{\prime }\right) \delta x\right) }  \label{aa3}
\end{equation}

We see that the edm induced phase shift at the output is exactly the same as
at the input to the 'amplifier'. Thus any measurement performed
subsequently, whether using a polarization analyzer or interferometer with
magnetic beam splitters or any other scheme, will yield the phase shift as
it was at the input and there is no amplification. Any attempts to measure
the increased displacement directly as a displacement are easily seen to be
completely impractical as is evidently recognized by the authors of \cite
{ringo}.

In the next section we show that a correct semi-classical argument leads to
the same conclusion.

\bigskip In response to the circulation of an earlier version of this paper, 
\cite{NIM}, Peshkin \cite{peshkin} introduces what he calls the ''No-Go''
theorem. Since $\psi _{\pm }\left( \overrightarrow{x},t\right) $ satisfy the
same Schroedinger equation in the amplifier region (the Hamiltonian is spin
independent)\ the quantity, $I\left( t\right) $, (see equ. (\ref{new5})) 
\begin{equation}
I\left( t\right) =\int d^{3}x\left[ \psi _{+}^{\ast }\left( \overrightarrow{x%
},t\right) \psi _{-}\left( \overrightarrow{x},t\right) \right]  \label{new6}
\end{equation}
is independent of time due to unitarity.

However this leaves open the possibility that the integrand over some
limited region of space might be time dependent, thus {\em allowing } the
amplifier to work without violating the theorem. Thus Peshkin states that
measuring the polarization ''in a small range of $x$ ..instead of over all $%
x $ at one time , to avoid the integral over all space in $I\left( t\right) $
..avoids the No-Go theorem in principle''. He also claims that the No-Go
theorem does not address interferometer experiments as ''there the phase
shift shows up as an overlap integral between two partial wave packets in
one emergent beam only, not as the conserved integral over all space. For
the same reason, the theorem also does not speak usefully to versions of the
proposal \cite{ringo} in which the phase shift is measured with a
Mach-Zender interferometer instead of a polarimeter''. The same argument is
presented in a recent Letter to the Editor of Nuc. Instr. and Meth. in
Physics Research, by Dombeck and Ringo, \cite{NIMDR}. The idea is that at
the output of an interferometer (with the spin states separated by a
magnetic mirror and travelling through the different arms of the
interferometer) the output (considering that one spin state is flipped
inside the interferometer) would be given by 
\[
\left| \psi _{+}\left( \overrightarrow{x},t\right) +\psi _{-}\left( 
\overrightarrow{x},t\right) \right| ^{2} 
\]
the cross terms giving the integral $I(t),$ equ. (\ref{new6}) but with the
integral taken only over one of the emergent beams. However the partial
beams $\psi _{\pm }\left( \overrightarrow{x},t\right) $ at the output \ of
the interferometer share all the same phase properties of the beams at the
input of the interferometer and the theorem will apply equally to both
situations. Be that as it may we propose a more stringent version of the
'No-Go' theorem, the 'Never-Go' theorem, i.e. the motion of the spin is
unaffected by a gravitational field due to the spin independence of the
Hamiltonian as we have shown above (equs. \ref{aaa5} and \ref{aa3})

We have shown that while an amplification of the 'displacement' betwen the
two spin states does occur in the 'gravitational amplifer' due to the
wavelength dependence of the index of refraction, this is virtually
unobservable; the phase shift between the two \ spin states remains
unchanged on traversing the region as does the direction of the neutron
polarization.

As we have not integrated over space or time our result shows that attempts
to avoid the No-Go theorem by confining the measurements to limited regions
will not work. The inapplicability of the theorem is not a sufficient
condition for the existence of an amplification effect.

\subsection{\protect\bigskip Semi-classical model}

This model employs the geometrical optics approach to quantum mechanics,
where each spin eigenstate is represented by a different trajectory. In a
sense this is the most difficult model as it is prone to misunderstanding.
In [1], at least three errors are made in application of the semi-classical
model to the spin amplification process. Let us first consider amplification
by vertical displacements, as discussed before equ. (3) above.

First, assuming that the concept of a displacement of the two spin
eigenstates is correct, we can calculate the gravitational effect on the
spin precession angle. The first error in [1] occurs when the assumption
that the phase between the two wave functions is simply the phase difference
between the two eigenstates evaluated at the respective maxima of the wave
function envelopes. This is a quantum mechanically incorrect procedure
because the phase determination does not commute with the determination of
the wave function center at a given time; in other words, it makes no more
sense to compare the phases between the two eigenfunctions at two distinct
spatial points than it does to compare the phases at two different times.

The correct procedure is to make a point by point comparison between the two
wave functions, then average over the two envelopes. This is the procedure
normally used when calculating the interference between two scalar or vector
fields as is commonly done in electrodynamics (see, for example, \cite{bandw}%
, Sec. 7.2). That this is the correct procedure can also be seen from the
fact that detection occurs at a single point in space-time (for example, the
neutron is absorbed on a $^{3}$He nucleus thereby ``collapsing'' the wave
function to a single space-time point; the spin direction is given by the
two wave function phases at that point in space-time).

We can now properly calculate the phase difference between the two
eigenstates at a fixed point in the final polarimeter/detector; it is the
phase difference between two classical trajectories that meet at the same
space-time point, initially separated a distance $\delta x$ perpendicular to
the momentum $\vec{k}$. The change in phase is simply the change in action
along the two trajectories, and this can be easily calculated to first order
by use of a theorem, which is of crucial importance to interferometry (but
universally ignored) due to Chiu and Stodolsky \cite{sandc}. This theorem
states that a change in the action when one of the endpoints of a classical
trajectory is displaced is given by 
\begin{equation}
\delta S=P_{\mu }^{D}\delta x_{\mu }^{D}-P_{\mu }^{0}\delta x_{\mu }^{0}
\end{equation}
with summation notation over spatial coordinates implied ($\mu =x,y,z$), and
where $S$ is the action (equal to the quantum mechanical phase up to a
factor of $\hbar $), $P_{\mu }$ refers to the momentum, $x_{\mu }$ the path
endpoint coordinates, $0$ refers to the trajectory beginning, and $D$ the
trajectory end (at the polarization analyzer/detector). As discussed
already, $\delta x_{\mu }^{D}$, that is, the relative displacement of the
path at the detector, is identically zero because a neutron is detected at a
single point (e.g., a polarized $^{3}$He nucleus). The displacement of the
path starting point is given by $\delta x_{x}\equiv \delta x$ as given by
equ. (2) above. However, it is assumed in [1] that $\delta x$ is
perpendicular to the neutron momentum; therefore, the change in action is
exactly zero, as given by the Chiu and Stodolsky theorem, and there is no
gravitational acceleration effect, in essence, by definition. Any other
effects that could change $S$, particularly those relating to the electric
field gradient or gravitational acceleration in the specific geometry given
in [1], enter only in second or higher order.

The above arguments can be immediately applied to the curved mirror
amplifier. We again assume that the trajectory endpoints must meet in order
for there to be an interference, and we calculated the change in action as
above. Again we find that $\delta S$ is identically zero to first order in
the electric field gradient.

The final semiclassical misconception in [1] concerns the use of an electric
field gradient over the storage volume. It seems to us that a larger (at
least two times) $\delta x$ (hence phase shift) can be generated by sending
the ``bipolarized'' neutrons from a region of zero electric field to a
region of constant high electric field. Each eigenfunction will acquire a
change in energy, hence a change in velocity, as it enters the electric
field region; although $\delta x$ only increases linearly in time, for a
given storage time $T_{1}$, suddenly accelerating the two eigenstates to
their final velocities would lead to a larger $\delta x$ than if the two
eigenstates were subjected to a weaker electric field gradient averaged over
the storage volume, but giving the same final velocities only after storing
for a time $T_{1}$. The implication is that the use of an electric field
gradient is completely pointless and only leads to a dilution of a possible
edm effect. However, this is not surprising based on the foregoing
considerations: to achieve the maximum sensitivity to an edm effect, whether
that be interpreted as $\delta x$ or a change in phase between the two spin
eigenfunctions, the interaction energy given by the usual Hamiltonian 
\begin{equation}
H=-\mu _{e}\vec{j}\cdot \vec{E}
\end{equation}
must be as large as possible over the duration of an experiment, for it is
the integral of $H$ over time that gives the relative action between the two
eigenstates. We thus see immediately that if the average magnitude of $E$ is
compromised by the wasting of electric field strength toward the
establishment of gradients in the system, the final net sensitivity to the
edm interaction given by $H$ must also be compromised.

\bigskip

\bigskip

\bigskip

\section{Conclusion}

We have analyzed the proposal for a new type of neutron edm experiment from
three different perspectives, and in each case, have arrived at the same
conclusion: The new technique offers no gain in sensitivity as compared to
the usual magnetic resonance technique. In fact, a careful analysis reveals
that the new method is inferior to the conventional methods. In [1], a
semiclassical approach was improperly used to analyze the proposed
technique, and a comparison with the atomic interferometer of Kasevich and
Chu \cite{kandc} was used to justify the approach. However, there really is
no point of comparison between the atomic interferometer and the system
described in [1]. The Kasevich and Chu ``interferometer'' is based on a
superposition of internal quantum states of the Cs atom, specifically, the
ground state hyperfine levels; there is no discussion here of the center of
mass wave function, but only of the phase difference between the internal
states. This phase difference evolves at the hyperfine frequency
(approximately 10 GHz) and the beauty of the system is that the freely
falling atom experiences a Doppler shift relative to an oscillator fixed in
the laboratory; this relative frequency shift in the accelerating system
makes possible, for example, a precise measurement of the Earth's
gravitational field. In a certain sense, the Kasevich and Chu system really
isn't an interferometer (this point is addressed in \cite{skl}); the
evolving internal quantum state phase difference serves as a clock which can
be compared to the stationary laboratory oscillator, and the system is best
described by the classical approach given above. We might invoke the quantum
or semiclassical model to calculate the result of some force that would
cause the two hyperfine levels to spatially separate; the result of this
calculation would simply show a diminishing of the internal interference
effect because the two hyperfine eigenfunctions no longer fully overlap in
space-time; in the limit where the separation is greater than the center of
mass wave function coherence length, the concept of a superposition of
internal quantum states entirely loses its meaning.

\section{\protect\bigskip Acknowledgment}

We would like to thank L. Stodolsky, who independently arrived at the same
conclusions, for sharing his thoughts on this matter with us, and for
encouraging us to write this manuscript.


\begin{references}
\bibitem{ringo}  Freedman, M.S., Ringo, G.R. and Dombeck, T.W., Nuc. Instr.
and Meth. in Physics Research {\bf A396}, 181 (1997).

\bibitem{NIM}  Lamoreaux, S.K. and Golub, R., Nuc. Instr. and Meth. in
Physics Research, to be published, 1999 and Los Alamos archive (xxx)
nucl-ex/9901007 v3

\bibitem{peshkin}  Peshkin, M., Los Alamos archive (xxx) nucl-ex/9903012 v2

\bibitem{NIMDR}  Dombeck, T.W. and Ringo, G.R., Nuc. Instr. and Meth. in
Physics Research, to be published, 1999

\bibitem{LG}  Golub, R. and Lamoreaux, S.K., Physics Reports {\bf 237},1,
(1994).

\bibitem{stnse}  G\"{a}hler, R., Golub, \ R. Habicht, K., Keller, T.,
Felber, J., Physica {\bf B229}, 1, (1996).

\bibitem{plane}  Golub, R., G\"{a}hler, R. and Keller, T. Am. J. Phys. {\bf %
62}, 779 (1994).

\bibitem{berry}  Berry, M., Lecture held at Technical University Berlin, 14
May, 1998.

\bibitem{bandw}  Wolf, E. and Born, M., {\it Principles of Optics, 6th ed.%
{\rm (Pergamon, Oxford 1980). }}

\bibitem{sandc}  Chiu, C. and Stodolsky, S., Phys. Rev. D {\bf 22}, 1337
(1980).

\bibitem{kandc}  Kasevich, M. and Chu, S., Phys. Rev. Lett. {\bf 67}, 181
(1991).

\bibitem{skl}  Lamoreaux, S.K., Int. Jour. of Mod. Phys. A {\bf 7}, 6691
(1992).
\end{references}
\end{document}